\newcommand{\wb}{\omega_{\mathrm{b}}}
\newcommand{\wt}{\wb\, t}

\newcommand{\ii}{\mathrm{i}}
\newcommand{\ee}{\varepsilon}

\newcommand{\fracc}[2]{\frac{\displaystyle #1}{\displaystyle #2}}

\newcommand{\N}{\mathcal{N}}
\renewcommand{\P}{\mathcal{P}}

\newcommand{\s}{\sigma}
\newcommand{\la}{\langle}
\newcommand{\ra}{\rangle}
\newtheorem{theorem}{Theorem}
\newcommand{\E}{\mathcal{E}}
\newcommand{\F}{\mathcal{F}}
\documentclass{elsart}
\usepackage{graphicx}
\usepackage{amsmath}
\usepackage{amssymb}
\journal{Physica D}

\begin{document}

\begin{frontmatter}
\author{JFR Archilla\thanksref{archilla}},
\author{J Cuevas}, 
\author{B S\'anchez--Rey} and 
\author{A Alvarez*} 
\address{Nonlinear Physics Group of the University of Sevilla
\\ Dep. F\'{\i}sica Aplicada I,
 ETSI Inform\'atica and  *Facultad de F\'{\i}sica, p. 5\\
  Avda Reina Mercedes s/n, 41012 Sevilla, Spain}

\thanks[archilla]{Corresponding author. E-mail: \texttt{archilla@us.es}}

\date{9 February 2003}
\title{Demonstration of the stability or instability of
multibreathers at low coupling}
\maketitle
\begin{abstract}
Whereas there exists a mathematical proof for one--site breathers
stability, and an unpublished one for two--sites breathers, the methods
for determining the stability properties of multibreathers rely in
numerical computation of the Floquet multipliers or in the weak
nonlinearity approximation leading to discrete non--linear Schr\"odinger
equations. Here we present a set of multibreather stability theorems
(MST) that provides with a simple method to determine multibreathers
stability in Klein--Gordon systems. These theorems are based in the
application of degenerate
perturbation theory to Aubry's band theory. We illustrate them with
several examples.
\end{abstract}

\end{frontmatter}
\section{Introduction}
\label{sec:introduction}
Discrete breathers are time--periodic, localized oscillations in
discrete systems due to a combination of nonlinearity and discreteness.
They have become a well understood phenomenon since the publication of
the proof of existence in Ref.~\cite{MA94}. This proof is the origin of
exact and powerful numerical methods to calculate them and to determine
their stability \cite{MA95}. A deeper insight has been achieved since
the introduction of Aubry's band theory \cite{A97}. The last reference,
together with Ref.~\cite{FW98} can be considered as reviews, although a
new one is badly needed after the huge development of the subject in the
last years.

The stability of one--site breathers is proofed in Refs.~\cite{MA95,A97}
under rather general conditions, the stability of the possible
two--sites breathers in analyzed in an unpublished theorem in
Ref.~\cite{M97}, page 69. Other methods use the hypothesis of small
amplitude oscillations or the rotating wave approximation
\cite{MJKA02,DDP97} leading to the Nonlinear Schr\"odinger equation
(DNLS), which is the actual equation analyzed, and, there are, of
course, the efficient but slow numerical methods mentioned above.

Here we propose a method based in the properties of the bands in Aubry's
bands theory and perturbation theory, to obtain the stability properties
of any multibreather at low coupling. In many cases it involves only
some modification of the coupling matrix and the knowledge of the
hardness/softness of the on--site potential. We will briefly summarize
the band theory in the next section, while we introduce the notation and
some basic concepts, on which our theory is based. However, we refer to
Ref.~\cite{A97,MA94,M97} for detailed explanations, as it will be long
and repetitive to expose it in detail.

In section~\ref{sec:multi} we develop the method for symmetric on--site
potentials or in--phase multibreathers, which is synthesized in a
theorem, commented in section~\ref{sec:comments}. Its scope is enlarged
to non--symmetric potentials in section~\ref{sec:multi}, and to
generalized Klein--Gordon systems in section~\ref{sec:general}. The
method is applied to several interesting examples in sections~\ref{sec:applications}
and \ref{sec:dark}. In Appendix~\ref{sec:gammamorse} we calculate
the value of a magnitude $\gamma$ used in our theory, and in 
Appendix~\ref{sec:curvature} we relate the curvature of the bands with
the characteristics of the on site potential.

\section{Band theory and notation}
\label{sec:notation}
\subsection{The Newton operator}
We consider Klein--Gordon systems with linear coupling described
by dynamical equations of the form:
\begin{equation}
\ddot{u}_n \,+\,V'(u_n)\,+\,\ee \,\sum_{m=1}^N
\,C_{n\,m}u_n\,=\,0\,\quad n=1,\dots,N \label{eq:klein}
\end{equation}
where the variables $u_n$ are functions of time $t$, $V(u_n)$ is
an homogeneous on--site potential, $V'$ its derivative,
$\dot{u}_n$ denotes derivation with respect to time, $N$ is the
number of oscillators, $C$ is a coupling constant matrix, which
can describe nearest neighbor or long--range interaction, and
includes the boundary conditions, and $\ee$ is the coupling
parameter. 
We use a notation
similar to quantum mechanics, i.e., $|u\ra \,\equiv [u_1(t),
\dots,u_N(t)]^\dag$ ($^\dag$ meaning the transpose matrix). 
Defining $V(u)=[V(u_1),\dots,V(u_N)]^\dag$ and
analogously its derivatives, equation (\ref{eq:klein})  can be
written as:
\begin{equation}
|\ddot{u}\ra  \,+\,|V'(u)\ra \,+\,\ee \,C \,|u\ra \,=\,0\,.
 \label{eq:kleinket}
\end{equation}
We suppose that the functions $u_n(t)$ are time--periodic (with period
$T$ and frequency $\wb$), time--reversible solutions and, therefore,
they can be written as cosine Fourier series with real coefficients
(some of these assumptions will be relaxed later). The (linear)
stability properties of a given solution $|u\ra $ depend on the
properties of the characteristic equation for the Newton operator
$\N_\ee$ given by
\begin{equation}
 \N_\ee(u)\, |\xi \ra  \,\equiv \,|\ddot{\xi} \ra \,
+\,V''(u)\cdot |\xi\ra \,+\,\ee\,C\,|\xi\ra =\,E\,|\xi\ra
\,,
 \label{eq:newton}
\end{equation}
where $\cdot$~product is the list product, i.e., $f(u)\cdot |\xi\ra $ is the
column matrix with elements $f(u_n(t))\,\xi_n(t)$. If $E=0$, this equation describes
the evolution of small perturbations $|\xi\ra $ of $|u\ra $. 

\subsection{The Floquet matrix}
Any solutions of Eq.~(\ref{eq:newton}) can be determined by the column
matrix of the initial conditions for positions and momenta 
$\Omega(0)=[\xi_1(0),\dots \xi_N(0),\pi_1(0),\dots \pi_N(0)]^\dag$, with
$\pi_n(t)=\dot{\xi}_n(t)$, and $\Omega(t)$ describes its evolution in the
space of coordinates and momenta. A base of solutions is given by the $2\,N$ functions 
with initial conditions $\Omega^\nu(0)$, $\nu=1,\dots,2\,N$,
with $\Omega^\nu_l(0)=\delta_{\nu\,l}$. We will identify often a given solution 
$\xi(t)$
of the Newton equation with the corresponding matrix of initial conditions
$\Omega(0)$. 
 As the Newton operator depends on
the $T$--periodic solution $u(t)$, it is also periodic, and the evolution
of the solutions of Eq.~(\ref{eq:newton}) can be studied by means of 
the Floquet operator, which maps $\xi(t)$ into $\xi(t+T)$. In a finite system 
this is equivalent to the Floquet  matrix $\F_E$ being given by:
\begin{equation}
\left[ \begin{array}{c} \{\xi_n(T)\} \\ \{ \pi_n(T) \} \end{array}  \right]=
\F_E\,\left[ \begin{array}{c} \{\xi_n(0)\}\\ \{ \pi_n(0)   \} \end{array} \right]\,
\end{equation}
These matrices can be easily constructed numerically, by integrating Eq.~(\ref{eq:newton})
$2\,N$ times from $t=0$ to $t=T$, with initial conditions $\Omega^\nu(0)$.
 Then, the $\nu$ column of $\F_E$ is given by $\Omega^\nu(T)$.

\subsection{Stability and bifurcations}
The eigenvalues $\{\lambda_l\}_{l=1}^{2\,N}$  of  $F_0$, called the Floquet multipliers,
determine the linear 
stability of the solution $u(t)$. If there is any eigenvalue with $|\lambda_l|>1$, 
the corresponding eigenfunction $\xi^l(t)$ grows with time and
 $u(t)$
is unstable; if $\lambda_l=1$, $\xi^l(t)$ is $T$-periodic;
if $\lambda_l=-1$, $\xi^l(t)$ is $2\,T$--periodic.
 However, $F_0$ (and any $F_E$) is a symplectic matrix, as it is 
derived from the symplectic system Eq.~(\ref{eq:newton}), which implies that if
$\lambda_l$ is a non--zero eigenvalue, so it is $1/\lambda$. $F_E$ is also real,
which in turn implies that $\lambda_l^*$ and $1/\lambda^*$ are also multipliers.
Therefore, the eigenvalues of $F_E$ come in groups of four, if they are not real, or
in pairs, if they are real or if $|\lambda_l|=1$. We can write the Floquet 
multipliers as $\{\exp(\ii \theta_l)\}$, with $\theta_l$, in general,
complex numbers. The complex number $\ii\theta_l$ are called the Floquet exponents, 
and $\theta_l$ the Floquet arguments.

The only possibility for the stability of $u(t)$ 
is  that all the eigenvalues of $F_0$ have moduli $1$, i.e., they are at the unit circle.
Therefore, the condition for linear stability of $u(t)$ can be described as
all the Floquet arguments for $F_0$ being real. Moreover, if $u(t)$ is stable and, then,
every $|\lambda_l|=1$, the Floquet multipliers come in
complex conjugate pairs with arguments $(\pm \theta_l)$ or as double $1$ or $-1$ 
($\theta=0$ or $\theta=\pm\pi$). If a parameter like the coupling $\ee$ is changed 
the multipliers of $F_0$ change continuously. Therefore, a bifurcation to an unstable solution
can only take place in three different forms: a)~Harmonic instability: 
two complex eigenvalues moving along the unit circle collide at $\lambda=1$ ($\theta=0$);
b)~Subharmonic instability: two complex eigenvalues moving along the unit circle 
collide at $\lambda=-1$ ($\theta=\pm\pi$); c)~Oscillatory or Hopf instability: 
two pairs of complex eigenvalues moving
along the unit circle collide at $\pm \theta \neq0$ and abandon the unit circle
as a quadruplet $(\lambda,1/\lambda,\lambda^*,1/\lambda^*)$. 

A Floquet multiplier of $F_0$ is always known. Calculating the derivative of 
Eq.~(\ref{eq:kleinket}) with respect to time we obtain that $\N \,\dot{u}=0$.
$\dot{u}$ is also $T$--periodic, therefore, it is an eigenfunction of $F_0$ with
multiplier $1$, and as they come in pairs, there is always a double $1$ multiplier.
$\dot{u}$ is called the {\em phase mode} because its meaning is that if 
$\dot{u}(t)$
is a solution of Eq.~(\ref{eq:kleinket}), $\dot{u}(t+dt)$ is also a solution.   
While the solution $u(t)$ exists, this double eigenvalue is always there.  Due to
the possible forms of the bifurcations, if all
the multipliers are isolated except the double $1$,
 the system is structurally stable, i.e., there is
a neighborhood in the space of the parameters where $u(t)$ is stable.
\subsection{Aubry's band theory}
However, it turns out that much information can be obtained by studying also 
the Floquet 
arguments for $E\neq 0$, with is known as Aubry's band theory~\cite{A97}.
 The set of points $(\theta,E)$, with $\theta$ a real Floquet   
argument of $F_E$ have a band structure. The fact that the Floquet multipliers
come in pairs of complex conjugate pairs brings about that if $(\theta,E)$ belongs
to a band, $(-\theta,E)$ does it too, i.e., the bands are symmetric with respect to 
$\theta$ and $\d E/\d \theta(0)=0$. As $\dot{u}$ has eigenvalue $1$ ($\theta=0$), there is always a band tangent
to the axis $E=0$ at $\theta=0$.
 There are at most $2\,N$ points for
a given value of $E$ and, therefore, 
there are at most $2\,N$ bands crossing any horizontal axes in
the space of coordinates $(\theta,E)$. The condition for linear stability of $u(t)$
is equivalent to the existence of $2\,N$ bands crossing the axis $E=0$ (including tangent
points with their multiplicity). If a parameter like the coupling $\ee$ changes,
the bands evolve continuously, and they can loss crossing points with $E=0$, 
bringing about the instability. 

We know more about the eigenfunctions of the Newton operator. The
fact that it is periodic, or, in other words, that it commutes with the
operator $\P$ of translation in time a period, $\P f(t)=f(t+T)$, implies 
that its eigenfunctions can be chosen simultaneously as
 eigenfunctions of $\P$, which by 
Bloch theorem, are given by $\xi(\theta,t)=\chi(\theta,t) \, \exp(\ii \,\theta \, t/T)$,
 $\chi(\theta,t)$
being a column matrix of $T$--periodic functions. It is straightforward
to check that $\exp (\ii\theta )$ is the corresponding Floquet multiplier,
and $\theta$ its Floquet argument ($F_E$ is simply the representation of
$\P$ in the base $\Omega^l$).
\subsection{Bands at the anticontinuous limit}
The key concept on the demonstration of breather existence~\cite{MA94}, single 
breather stability~\cite{A97} and the present paper is the 
{\em anticontinuous limit}, i.e., the system with all the oscillators 
uncoupled, $\ee=0$, in Eq.~(\ref{eq:klein}--\ref{eq:kleinket}). 
At the anticontinuous limit, Eq.~(\ref{eq:klein})
reduces to $N$ identical equations:
\begin{equation}
 \ddot{u}_n \,+\,V'(u_n)\,=\,0\, .
\label{eq:dyn0}
 \end{equation}
Supposing that we consider time--reversible solutions around a single minimum of $V$,
there are only three different solutions: a)~oscillators at rest
$u_n\,=\,0$; b)~excited oscillators with identical $u_n(t)$, hereafter
denoted $u^0(t)$;
 c)~excited oscillators with a phase difference of
$\pi$ with the previous ones, given by $u_n(t)\,=\,u^0(t\,+\,T/2)$. Each
site index is given a code $\sigma_n$, which takes elements in
$\{0,1,-1\}$, where $\sigma_n\,=\,0$ represents an oscillator at rest,
$u_n\,=\,0$; $\sigma_n\,=\,1$, an oscillator with solution $u^0(t)$; and
$\sigma_n\,=\,-1$, the solution $u^0(t+T/2)$. The matrix of codes
$\sigma\,=\,[\sigma_1,\dots,\sigma_N]^\dag$ represents the state of the
system at the anticontinuous limit.

We suppose that there are $p$
oscillators at rest and $N\,-\,p$ excited oscillators.
Equation~(\ref{eq:newton}) at $\ee\,=\,0$ becomes
\begin{equation}
 \N_0(u)\, |\xi \ra  \,\equiv \,|\ddot{\xi}\ra  \,
+\,V''(u)\cdot |\xi\ra \,=\,E\,|\xi\ra \,,
 \label{eq:newton0}
\end{equation}
or, equivalently, $N\,-\,p$ identical equations:
\begin{equation}
\ddot{\xi}_n \,+\,V''(u_n)\,\xi_n=\,E\,\xi_n\,,
 \label{eq:newtonexcited}
\end{equation}
with only a periodic eigenfunction for $E\,=\,0$, the phase mode
$\dot{u}_n(t)$    for the isolated, excited oscillators. The other one
is the growth mode, (see Appendix~\ref{sec:curvature})
 which is not bounded and will not be used in this work.
  They give rise to $N-p$ bands tangent to the axis
$E\,=\,0$, shown in Fig.~\ref{fig:bands0}. They have positive
curvature at $(\theta,E)=(0,0)$ 
if the on--site potential is soft ($\d H/\d\omega_b<0$) and
negative if it is hard ($\d H/\d\omega_b>0$), as it is demonstrated in 
Appendix~\ref{sec:curvature}. The possibility of $\d H/\d\omega_b=0$ 
is excluded by the conditions of the breather existence theorem~\cite{MA94}, 
which means that the on-site potential is truly nonlinear
for the isolated oscillators with the breather frequency $\wb$. If, as we are supposing here,
 the on--site potentials $V(u_n)$
are identical, the solutions $u_n(t)$ are also identical except for a
change of phase and the bands are superposed, if not, we might have different
bands but with the same general shape.

The remaining $p$ equations corresponding to the oscillators at rest
are of the form:
\begin{equation}
\ddot{\xi}_n \,+\,(\omega_0)^2\xi_n=\,E\,\xi_n\,,
 \label{eq:newtonrest}
\end{equation}
with $\omega_0\,=\,\sqrt{V''(0)}$. They have only a $T$-periodic
solution for $E\,=\,0$ , the null solution, due to the non--resonance
condition for breather existence  
$n\omega_b\,\neq\,\omega_0$ ($p\in \mathbb N$)~\cite{MA94}, i.e.,
none of the harmonics of the breather resonates with the oscillators at rest.
They provide $p$ identical bands which are easily calculated solving the equation
above. Its solutions are $\xi=\exp(\pm\ii \sqrt{\omega_0^2-E}\,\,t)$,
 with Floquet multipliers
 $\exp(\pm\ii \sqrt{\omega_0^2-E}\,\,T)$ and 
 Floquet arguments $\theta=\pm \sqrt{\omega_0^2-E}\,\,T$. That is,
 the bands are given by $E=\omega_0^2-\wb^2 (\theta/T)^2$, where
 $\theta$ can be reduced to the first Brillouin zone $[-\pi,\pi]$ by the
 addition of $2\pi p$, $p\in\mathbb Z$.
 If the oscillators have different rest frequencies, the rest bands are not
 superposed but they  have the same characteristics.
  
  The rest bands are also shown in Fig.~\ref{fig:bands0}.
     Note that this figure is only a sketch, for clarity,
      as very often some excited oscillators bands are very
 flat and difficult to appreciate at the same scale. This sketch reproduces,
  however,
 the basics facts of the band structure. 
 
 When the coupling is switched on, the degeneracy of the bands is generically raised.
 All the tangent bands at $(\theta,E)=(0,0)$ except one, than continues there,
 move upwards or downwards.
 If the on--site potential is soft, Fig.~\ref{fig:bands0} (left), and a band moves upwards,
 a double tangent point with the axis $E=0$ is lost, or, in other terms, a pair of
 Floquet arguments of $F_0$ becomes complex and the solution $u(t)$ is unstable. 
 If the on site potential is hard, Fig.~\ref{fig:bands0} (right), the same occurs when a 
 band moves downwards. 

\section{Multibreathers stability}
\label{sec:multi}

In this work use degenerate perturbation theory \cite{G95} to demonstrate the
stability or instability of the breathers of any code.  
Degenerate
perturbation theory establishes that if $\N_0$ is a linear operator with a
degenerate eigenvalue $E_0$, with eigenvectors $\{|v_n\ra \}$, which are
ortonormal with respect to a scalar product, i.e., $\la v_n|v_m\ra
\,=\,\delta_{n\,m}$, and if $\ee\,\tilde{\N}$ is a perturbation of
$\N_0$, with $\ee$ small; then, to first order in $\ee$, the eigenvalues
of $\N_0\,+\,\ee\,\tilde{\N}$ are $E_0 \,+\ee\,\,\lambda_i$, with
$\lambda_i$ being the eigenvalues of the perturbation matrix $Q$ with
elements $Q_{n\,m}\,=\,\la v_n|\tilde{\N}|v_m\ra $. Note that
perturbation theory as described in the reference cited, is time
independent perturbation theory, but our time variable $t$ is their
spatial coordinate $x$.

Let us consider again Eq.~({\ref{eq:newton}) with zero coupling, $\ee=0$. 
As explained in the previous 
section, if there are $N-p$ excited oscillators, there
$N-p$ zero eigenvalues corresponding
 to the $T$--periodic eigenfunction $\dot{u}_n$ (the phase modes of the
 isolated oscillators),
 or in other words, a $N-p$
times degenerate eigenvalue $E_0=0$ if we restrict the domain of $\N$
to periodic functions. What we need to know is the sign
of this degenerate eigenvalues when the coupling is switched on.

To apply degenerate theory to multibreather stability we need to
identify the perturbation operator, a suitable scalar product and a
ortonormal basis of the eigenspace with eigenvalue $E_0=0$.
The scalar product is defined as:
\begin{equation}
\la \xi_1|\xi_2\ra
\,=\,\sum_{n=1}^{N}\int_{-T/2}^{T/2}\xi_1^*(t)\xi_2(t)\,. \d t\,.
\label{eq:scalar}
\end{equation}
Let us suppose initially that all the excited oscillator are identical
and vibrate in phase. We will denote by $u^0(t)$ these 
identical solutions. In this case, all the phase modes are also 
identical
and will be denoted as $\dot{u}^0$.
The $N-p$ elements of the basis are
\begin{equation}
 |n\ra \,=\,\frac{1}{\mu}[0,\dots,0,\dot{u}^0,0,\dots,0]^\dag,
 \label{eq:basisn}
\end{equation}
with the non-zero element at the position $n$, $n$ being the index
of the excited oscillators, and
$\mu\,=\,\sqrt{\int_{-T/2}^{T/2}(\dot{u}^0)^2\,\d t}$. It is
straightforward to check that they are ortonormal.

%%%%%%%%%%%%%%%%%%%%%%%%%figure1%%%%%%%%%%%%%%%%%%%%%%
\begin{figure}
\begin{center}
    \includegraphics[width=0.9\textwidth]{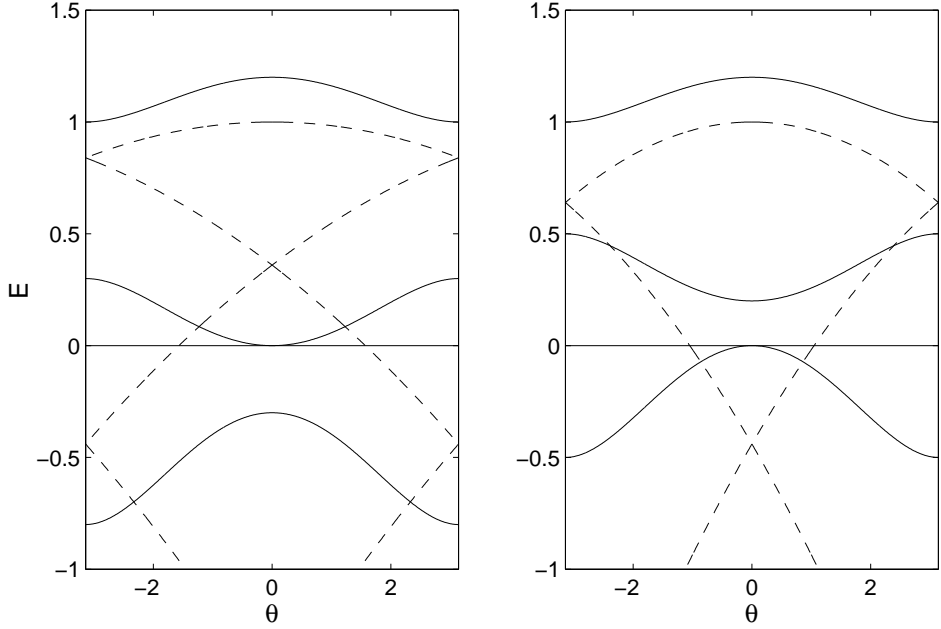}
\end{center}
\caption{Sketch of the band structure for a soft on--site
potential (left) and a hard on--site potential (right). The
continuous lines correspond to excited oscillators and the dashed
ones to oscillators at rest. } \label{fig:bands0}
\end{figure}

It is enough to know the eigenvalues corresponding to periodic and
real solutions, i.e., with Floquet argument $\theta\,=\,0$, as the
intersections of the bands with the axis $\theta\,=\,0$ correspond
to periodic solutions of (\ref{eq:newton}). Therefore the $N-p$
$|n\ra $'s form the basis of the degenerate eigenvalue $E_0=0$
needed to apply perturbation theory.

To obtain the perturbation operator, we expand in Taylor series 
Eq.~(\ref{eq:newton}) at $\ee\,=\,0$ and obtain to first order in $\ee$\;:
 \begin{eqnarray}
 \N_\ee(u)\, |\xi \ra
 \,&=& \,|\ddot{\xi}\ra \,+\,
 \,V''(u)\cdot|\xi\ra \,
+\,\ee\,\big(\,V'''(u)\cdot u_\ee\,\cdot|\xi\ra
\,+\,C\,|\xi\ra \,\big)\,=\nonumber \\
&=&\, (E_0\,+\,\ee\,\lambda_i)|\xi\ra \,,
 \end{eqnarray}
with $u_\ee=\big(\frac{\displaystyle \partial u}{\displaystyle \partial
\ee}\big)_{\ee=0}$ and $u$ is also the $\ee=0$ solution.

 Therefore the perturbation operator of $\N_0$ is
 \begin{equation}
 \tilde{\N}\, |\xi \ra  \,= V'''(u))\cdot u_\ee \cdot |\xi\ra \,+ \,
C\,|\xi\ra \,,
 \end{equation}
where $u$ and $u_\ee$ are calculated at $\ee=0$. We do not know $u_\ee$,
but deriving with respect to $\ee$, at $\ee\,=\,0$, the dynamical
equations (\ref{eq:kleinket}) we obtain:
\begin{equation}
|\ddot{u}_{\ee}\ra \,+\,V''(u)\cdot|u_{\ee}\ra \,+\,C\,|u\ra
\,=\,0\quad \mbox{or}\quad N_0|u_\ee\ra \,=\,-C\,|u\ra \,.
\label{eq:eque}
\end{equation}
 The perturbed eigenvalues $E_i$ are $\ee\,\lambda_i$\;,
$\lambda_i$ being the eigenvalues of the perturbation matrix $Q$, with
elements $\la n|\tilde{\N}|m\ra $. The matrix $\tilde{C}$ of elements
$\la n|C|m\ra $ is simply the matrix $C$ without the columns and rows
corresponding to the oscillators at rest. The other terms are:
\begin{eqnarray}
 \la n|V'''(u)\cdot u_\ee |m\ra \,=&& \nonumber \\
  \frac{1}{\mu^2}\,\int_{-T/2}^{T/2}
  [\dots,0, \dot{u}_n^0,0,\dots] \, &&
 [ \dots,0,V'''(u^0)\, u_{m, \ee}
\, \dot{u}_m^0,0,\dots]^\dag
 \, \d t \, = \nonumber \\
  \mbox{}&& \fracc{\delta_{n\,m}}{\mu^2}\,\int_{-T/2}^{T/2}\dot{u}^0
   \,V'''(u^0)\,u_{n,\ee}
 \, \dot{u}^0 \,\d t\,,
 \label{eq:integral}
\end{eqnarray}

with $u_{n,\ee}=\big( \frac{\displaystyle \partial u_n}{\displaystyle
\partial
\ee}\big)_{\ee=0}$. Thus, only the diagonal elements $ \la
n|V'''(u)\cdot u_\ee
\cdot |n\ra $ are non--zero. To calculate the last integral in
(\ref{eq:integral}) we will integrate by parts and use that the integral
in a period of the derivative of a periodic function is zero. Besides,
the functions $u_{n,\ee}$ are periodic as the coefficients of their
Fourier series are given by the derivatives with respect to $\ee$ of the
Fourier coefficients of $u_n$. In the deduction below, all the integral
limits are $-T/2$ and $T/2$, and the terms between brackets from
integration by parts will be zero. The last integral in
Eq.~(\ref{eq:integral}) becomes:
\begin{eqnarray}
\left[\dot{u}^0\,u_{n,\ee}\,V''(u^0)\right]_{-T/2}^{T/2} \,
-\,\int \, V''(u^0)\dot{u}^0\, \dot{u}_{n,\ee} \,\d t \,-\,
 \int\,V''(u^0)\,\ddot{u}^0\,u_{n,\ee}\,\d t \,&=&\nonumber
\\
 -\left[V'(u^0)\,\dot{u}_{n,\ee},\right]_{-T/2}^{T/2}
 \,+\,\int\,V'(u^0)
 \,\ddot{u}_{\ee,n},\d t\,-\int\,V''(u^0)\,\ddot{u}^0\,u_{n,\ee}\,\d t\,&=&\nonumber\\
 -\,\int\, \ddot{u}^0\,(\ddot{u}_{n,\ee} \,+\, V''(u^0)\,u_{n,\ee})\d t\mbox{}&\mbox{}&\mbox{}
\label{eq:larga}
\end{eqnarray}
The term between parentheses, is the $n$ component of the lhs of
Eq.~(\ref{eq:eque}), i.e., it becomes $-\sum_m C_{n\,m}\,u_m^0$, where
$u_m^0\,=\,u^0$, if the oscillator $m$ is excited, and zero otherwise,
i.e., it is $-\sum_m
\tilde{C}_{n\,m}\,u^0\,= \,
 -(\sum_m \tilde{C}_{n\,m})\,u^0$. Equation (\ref{eq:larga}) becomes:
 \begin{eqnarray}
(\sum_m
\tilde{C}_{n\,m})\,\int\,\ddot{u}^0\,u^0\,dt\,&=&\mbox{}\nonumber
\\(\sum_m \tilde{C}_{n\,m})\,\left(\left[
\dot{u}^0\,u^0\right]_{-T/2}^{T/2}-
 \int (\dot{u}^0)^2\,dt\right)\,&=&\,-(\sum_m
 \tilde{C}_{n\,m})\mu^2\,.
 \end{eqnarray}
That is, equation (\ref{eq:integral}), leads to:
\begin{equation}
\la n|V'''(u)\cdot u_\ee\cdot|n\ra \,=\,-\sum_m\tilde{C}_{n\,m}\quad \textrm{at}\quad \ee\,=\,0\,.
 \label{eq:diagonalw1}
\end{equation}
Therefore the diagonal elements of the perturbation matrix $Q$ are
\begin{equation}
Q_{n\,n}=\,-\,\sum_m\,\tilde{C}_{n\,m}\,+\,\tilde{C}_{n\,n}\,=\,
-\,\sum_{\forall\, m\neq n}\,\tilde{C}_{n\,m}\,=\,\sum_{\forall\, m\neq n}\,Q_{n\,m}\, .
\end{equation}
To summarize, the perturbation matrix $Q$ is given by:
\begin{equation}
Q_{n\,m}\,=\,\tilde{C}_{n\,m},\quad n\neq m\quad,\quad
Q_{n\,n}\,=\,-\,\sum_{\forall\, m\neq n}\, Q_{n\,m}\,,
\label{eq:matrixq}
\end{equation}
$\tilde{C}$ being the coupling matrix without the $p$ rows and
columns corresponding to oscillators at rest.

This result can be very easily extended, to the case when there
are oscillators out of phase, at least for an even potential
$V(u)$. In this case $u^0(t+T/2)\,=\,-u^0(t)$ and
$\dot{u}^0(t+T/2)\,=\,-\dot{u}^0(t)$.   Let us define a new code
$\tilde{\sigma}$, which is equal to $\sigma$, but with $+1$
instead of zeros, and perform the ansatz $u_n=\tilde{\sigma}_n
\,\tilde{u}_n$. By substitution in the dynamical equations
(\ref{eq:klein}), and taking into account that
$V'(u_n)\,=\,V'(\tilde{\sigma}_n \,\tilde{u}_n)\,=\,
\tilde{\sigma}_n\,V'(\tilde{u}_n)$, because $V'(u)$ is odd, we
obtain:
\begin{equation}
\tilde{\sigma}_n \,\ddot{\tilde{u}}_n
\,+\,\tilde{\sigma}_n\,V'(\tilde{u}_n)\,+\,
 \sum_m\,C_{n\,m}\,\tilde{\sigma}_m\,\tilde{u}_m\,=\,0\,.
\label{eq:withansatz}
\end{equation}
Multiplying by $\tilde{\sigma}_n$, and using that its square is
$+1$ we get:
\begin{equation}
\ddot{\tilde{u}}_n \,+\,V'(\tilde{u}_n)\,+\,
 \sum_m\,\tilde{\sigma}_n\,C_{n\,m}\,\tilde{\sigma}_m\,\tilde{u}_m\,=\,0\,.
\label{eq:transformed}
\end{equation}
As the elements of $C$ with an index corresponding to  an
oscillator at rest will not appear in the calculations, we can
define a new coupling matrix, with elements
$\sigma_n\,C_{n\,m}\,\sigma_m$, i.e.,
$\textrm{Diag}(\sigma)\,C\,\textrm{Diag}(\sigma)$. The eigenvalues
$E_i$ will have the right sign, but the eigenfunctions
$\tilde{\xi}$ of the Newton operator corresponding to
(\ref{eq:transformed}) have to be transformed to recover the
original ones by $|\xi\ra
\,=\,\textrm{Diag}(\sigma)\,|\tilde{\xi}\ra $

The result can be summarized in the following way:

\begin{theorem}[Symmetric MST] \label{th:MST} Consider a
 Klein--Gordon system, with homogeneous on--site potential $V$, linear
coupling matrix $C$ and coupling parameter $\ee$, and a multibreather
given by a vector code $\sigma$ at zero coupling, with $p$ zero
elements, and $N\,-\,p$ non--zero elements. The potential $V$ must be
even if there is any $-1$ in $\sigma$. We define the following matrices:
$S\,=\,\textrm{Diag}(\sigma)$; $\tilde{C}$ the $N\,-\,p$, squared matrix
equal to $S\,C\,S$ but without the rows and columns corresponding to the
zero codes in $\sigma$; $Q$, the perturbation matrix, with non--diagonal
elements $Q_{n\,m}=\tilde{C}_{n\,m}$, and diagonal elements
$Q_{n\,n}\,=\,-\sum_{\forall\, m\neq n}\tilde{C}_{n\,m}$. Let
$\{\lambda_i\}_{i=1}^{N-p}$ be the eigenvalues of $Q$ and suppose that
there is only one zero among them. Then:\\
 a)~if $V(u_n)$ is hard and there is any negative value in
 $\{\ee\,\lambda_i\}$ the multibreather at low coupling will be
 unstable, and stable otherwise.\\
 b)~if $V(u_n)$ is soft and there is any positive value in
 $\{\ee\,\lambda_i\}$ the multibreather at low coupling will be
 unstable, and stable otherwise.
\end{theorem}
There is always a zero
eigenvalue is explained below. If there are more than one the stability
is undefined within first order perturbation theory.

\section{Comments}
\label{sec:comments}

\subsection{Global phase mode}
\label{sec:global}
 The perturbation matrix $Q$ has always a zero eigenvalue
corresponding to the global phase mode. It can be useful to make it
apparent as it reduces the order of the secular equation by an unity.
Considering the matrix $Z=Q-\lambda\,I$, $\det(Z)=0$ is the secular
equation. Its roots are the same if we change $Z$ to $Z'$ obtained by
summing to the first column the rest of them, i.e.,
$Z'_{n\,1}\,=-\sum_{\forall
\,m\neq n}\tilde{C}_{n,m}
\,-\lambda\,+ \,\sum_{\forall\,m\neq n}\tilde{C}_{n,m}\,=\,-\lambda$. Now, subtracting
the first row from the rest of them, the first column has only a non
zero element, which is equal to $\-\lambda$. Therefore, $\lambda=0$ is
an eigenvalue of $Q$ and the secular equation is reduced by an unity.

The corresponding eigenvector $\tilde{x}$ satisfies
$Q\,\tilde{x}\,=\,0$, i.e., $-\,(\sum_{\forall\,m\neq
n}\,\tilde{C}_{n\,m}\,)\,\tilde{x}_n\,+ \,\sum_{\forall\, m\neq n}
\,\tilde{C}_{n\,m}\,\tilde{x}_m\,=\,0$, $\forall n$, with solution
$\tilde{x}_n\,=\,1$, $\forall n$. Therefore,
$x\,=\,S\,\tilde{x}\,=\sigma$, which corresponds to the global phase
mode.
\subsection{Extended and reduced forms}
\label{sec:extended}   Note that the code $\sigma$ is the matrix
of components of the multibreather at zero coupling in a N--dimensional
basis, $\{\,|n\ra\} $, with the only non zero element $u^0$ at the
position $n$, including the indexes of all the oscillators. For
numerical calculation and interpretation of the eigenvalues and
eigenvectors of $Q$ it can be more convenient to define
$\tilde{C}\,=\,S\,C\,S$, i.e, conserving the rows and columns
corresponding to the rest oscillators but all their elements set to
zero. It will be denoted the {\em extended} form while the previous one
is the {\em reduced} form. The price for using the extended form is to
obtain $p$ extra zero eigenvalues, which causes no harm, except if there
are more than $p+1$, as it can make more difficult the interpretation of
the eigenvectors.
\subsection{Diagonal elements}
\label{sec:diagonal}
  The diagonal elements of
$C$ play no role, because they do not appear in the coupling matrix.
Therefore a coupling of the spring--type as $\ee\,(\,2\,u_n -\,u_{n-1}
-\,u_{n+1})$ gives identical results as the dipole-type
$-\ee\,(u_{n-1} +u_{n+1})$.
\subsection{2D and 3D lattices}
Note that the theorem applies equally to 2D and 3D lattices,
provided that we can number the sites with an index $n$, or
consider $n$ as a multi--index, and construct the coupling matrix
$C$.
\section{Non--symmetric potentials}
 \label{sec:non--symmetric}
 The extension to non--symmetric potentials $V(u_n)$ is
straightforward, but it involves some numerical or analytical
calculation. That is the reason for not including it in
theorem~\ref{th:MST}, and we present it here as another theorem instead.
If the potential is non--symmetric, there are two time--symmetric,
$T$-periodic solutions to the equations for the isolated oscillators at
zero coupling, Eq.~(\ref{eq:dyn0}): $u^0(t)$, with code $+1$, and
$u^0(t+T/2)$, which is no longer $-u^0(t)$, with code $-1$. The $N-p$
elements of the basis, $|n\ra $ will be as in Eq.~(\ref{eq:basisn}) but
with $\dot{u}^0(t+T/2)$ instead of $\dot{u}^0(t)$ if $\sigma_n\,=\,-1$.
  Let us define the symmetry coefficient, $\gamma$, in the following way:
\begin{equation}
\gamma\,=\,-\,\frac{\int_{-T/2}^{T/2}\dot{u}^0(t)\,\dot{u}^0(t+T/2)\,\d
t} {\int_{-T/2}^{T/2}\dot{u}^0(t)\,\dot{u}^0(t)\,\d t}
\label{eq:gammau}
\end{equation}
If $\{z_k\}_{k=0}^{k_m}$ are the Fourier coefficients of a
$k_m$--truncated Fourier series of $u^0(t)$, i.e., $u^0(t)\,=\,
z_0\,+\,\sum_{k=1}^{k_m}2\,z_k\,\cos(\,k\,\omega_b \,t)$, then:
\begin{equation}
\gamma\,=\,-\,\frac{\sum_{k=1}^{k_m} (-1)^k \,k^2\,z_k^2}
{\sum_{k=1}^{k_m}\,k^2\,z_k^2}\,. \label{eq:gammaz}
\end{equation}
Note that $\gamma$ depends on the breather frequency through the
$\{z_k\}$, and it is positive and smaller than $1$, being its value $\gamma\,=\,1$
for the symmetric potential case.
  Following the deduction of theorem~\ref{th:MST} it is
  straightforward to obtain:

\begin{theorem}[Non--symmetric MST]
\label{th:non--symmetric}
 The generalization of theorem~\ref{th:MST} for non--symmetric potentials
consists of constructing the perturbation matrix $Q$ in the
following way:
\\a)~If $n$ or $m$ are indexes corresponding to oscillators at rest
$Q_{n\,m}\,=\,0$.
\\b)~If $n$ and $m$, $n\neq m$, are indexes corresponding
to excited oscillators in phase, i.e., $\s_n\,\s_m\,=\,1$,
$Q_{n\,m}\,=\,C_{n\,m}$.
\\c)~If $n$ and $m$, $n\neq m$, are indexes corresponding
to excited oscillators out of phase , i.e., $\s_n\,\s_m\,=\,-\,1$,
$Q_{n\,m}\,=-\gamma\,C_{n\,m}$.
\\d)~The diagonal elements are $Q_{n\,n}=-\sum_{\forall\, m\neq n}Q_{n\,m}$
\\e)~The rest of the procedure is exactly the same as described in
theorem~\ref{th:MST}, getting rid of the rest columns and rows (reduced
form) or not (extended form), calculating the eigenvalues
$\{\lambda_i\}_{1}^{N-p}$ of $Q$ and obtaining the stability properties
of the multibreather from the signs of the $\{\ee\lambda_i\}$ and the
hardness/softness of the on--site potential.
\end{theorem}

The coefficient $\gamma$ depends on the on--site potential and the
frequency $\wb$. We demonstrate in the Appendix~\ref{sec:gammamorse} that $\gamma=\wb$ for
the Morse potential. Figure~\ref{fig:gammas} shows some numerically
calculated values for different potentials. They tend to $+1$ when $\wb
\rightarrow \omega_0=1$ because at this limit
$\dot{u}^0(t)=A\,\sin(\wb\,t)$ and $\dot{u}^0(t+T/2)=-\dot{u}^0(t)$ as
in the symmetric on--site potential case.
 Generally speaking, if the frequencies are not very far away
from the linear frequency $\omega_0$, $\gamma$ is usually close enough
to the unity and the sign of the eigenvalues do not change with respect
to the symmetric on--site potential case $\gamma=1$.
%%%%%%%%%%%%%%%%%%%%%%figure2%%%%%%%%%%%%%%%%%%%
\begin{figure}
\begin{center}
        \includegraphics[width=\textwidth]{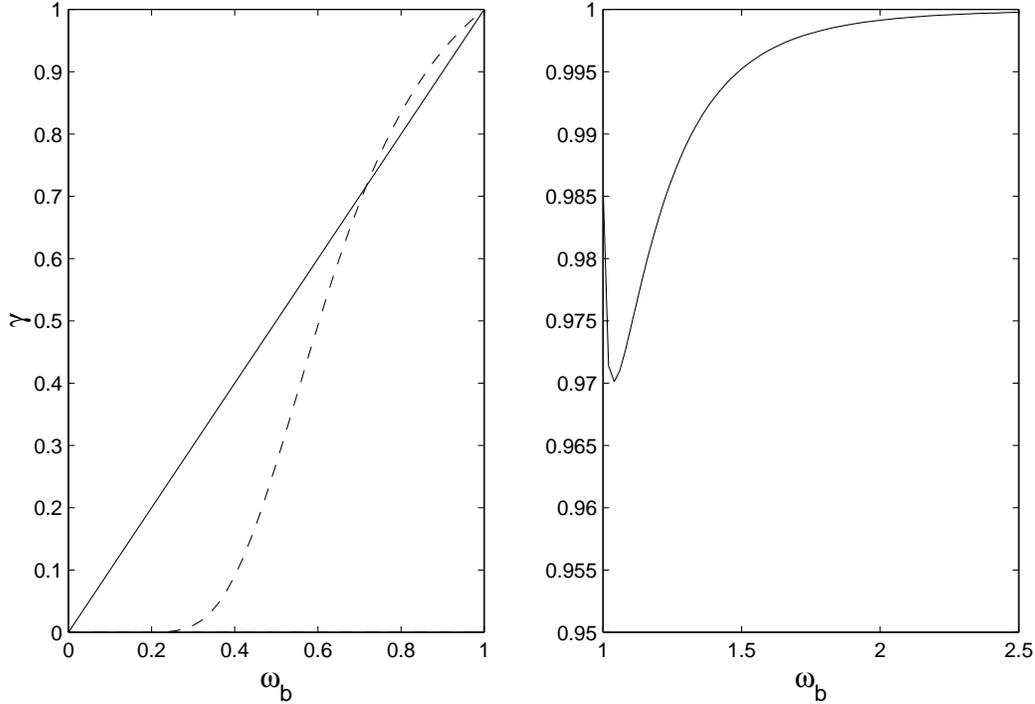}
\end{center}
\caption{Some values of the symmetry coefficient $\gamma$. Left,
soft on--site potentials: dashed line:
$V(u_n)=\frac{1}{2}u_n^2-\frac{1}{3}u_n^3$; continuous line:
$V(u_n)=\frac{1}{2}(\exp(-u_n)-1)^2$. Right, hard on--site potential:
$V(u_n)=\frac{1}{2}u_n^2+\frac{1}{3}u_n^3+\frac{1}{4}u_n^4$. Rest
frequency $\omega_0=1$.}\label{fig:gammas}
\end{figure}

\section{Generalization} \label{sec:general}
Theorems~\ref{th:MST} and \ref{th:non--symmetric} are equally applicable
if the system has inhomogeneities at the coupling matrix but not if they
are at the on--site potential. Here we introduce the generalization to a
broad class of Klein--Gordon systems, including different masses or
inertia momenta, inhomogeneous on-site potentials with several minima,
nonlinear coupling, solutions with different phases (i.e, not restricted
to time--reversible solutions). Let us consider a Klein--Gordon system
with Hamiltonian
\begin{equation}
H=\sum_n \left(\frac{1}{2}\,m_n\,\dot{u}_n^2\,+\,V_n(u_n)\right)
\,+\,\ee \,W(u)\, ,
 \label{eq:ghamiltonian}
\end{equation}
and dynamical equations
\begin{equation}
m_n\,\ddot{u}_n \,+\,V_n'(u_n)\,+\,\ee \,\frac{\partial
W(u)}{\partial {u_n}} \,=\,0\, ,
 \label{eq:gklein}
\end{equation}
which at zero coupling become:
\begin{equation}
m_n\,\ddot{u}_n \,+\,V_n'(u_n)\,=\,0\,\quad \forall n\,.
\label{eq:gdyn0}
\end{equation}
Suppose we can label the solutions of these equations, with given
frequency $\wb$ and period $T$, with a code composed of an index for the
site $n$, another index for the different minima and solutions
(biperiodic, etc, different potential wells) $\sigma_n$, which can be
different for each site. If we considering a periodic solution for the
$n$--th equation , it can be written as a $2\,\pi$ periodic function
$g_{n\, \wb \,
\sigma_n}(\wb\,t+\alpha)$. Let us suppose that $g_{n\, \wb \,
\sigma_n}$ is chosen such that $\alpha=0$ corresponds to the time
symmetric solution with
 $\dot{g}_{n\, \wb \,\sigma_n}(0)=0$ and
 $g_{n\, \wb \, \sigma_n}(0)>g_{n\,\wb\,\sigma_n}(\pi)$.
Therefore, the solutions of equations~(\ref{eq:gdyn0}) are determined by
the code $\{n,\sigma_n,\alpha_n\}$. If we consider a given solution at
the anticontinuous limit we can simply write it as
$u=[u_1^0,\dots,u_N^0]^\dag$, but now, the $u_n^0$ are, in general,
different, because of the particular multibreather solution $u$, whose
stability we are interested in.

The characteristic equations for the Newton operator $\N_\ee$ are given
by:
\begin{equation}
(\N_\ee\,\xi)_n \,\equiv\, m_n \,\ddot{\xi}_n \,+\,
V_n''(u_n)\,\xi_n\,+\,\ee\,\sum_m\frac{\partial^2 W (u)}{\partial
u_n\,\partial u_m}\,\xi_m\,=\,E\,\xi_n \,\label{eq:gnewtonee}
\end{equation}
which to first order in $\ee$ calculated at $\ee=0$ lead to
\begin{equation}
(\N_0 \,\xi)_n\, + \,\ee \,(\tilde{\N}\,\xi)_n \,=
\,(E_0\,+\,\ee\,\lambda)\,\xi_n\,, \label{eq:gnewton0newtondot}
\end{equation}
where:
\begin{eqnarray}
(\N_0 \,\xi)_n\,&
 \equiv &\,m_n\,\ddot{\xi}_n\,+\,V_n''(u^0_n)\,\xi_n\,=\,E_0\,\xi_n\,,
\label{eq:gnewton0}\\
 (\tilde{\N}\xi)_n\,
 &\equiv &  V'''(u_n^0)\,u_{n,\ee}\,\xi_n
 \,+\,\sum_m\,\frac{\partial^2 W(u^0) }{\partial u_n\,\partial
u_m}\,\xi_m\,=\,\lambda \,\xi_n\,.
 \label{eq:gnewtondot}
\end{eqnarray}
 with $u_{n,\ee}\,=\,\big(\frac{\displaystyle \partial
u_n}{\displaystyle \partial \ee}\big)_{\ee=0}$. The first set of
equations is the characteristic equation for the Newton operator
at zero coupling, and the second one for the perturbation operator
$\tilde{\N}$. The only periodic solution of each
equation~(\ref{eq:gnewton0}) with $E_0=0$ for a given solution
$u_0$ of (\ref{eq:gdyn0}) , is $\dot{u}_n^0$. Thus, we can
construct a ortonormal basis of $\ker(N_0)$, restricted to
periodic solutions (and, therefore, that can be chosen real) of
given period $T$, as $| n \ra
\,=\,\frac{1}{\mu_n}[0,\dots,0,\dot{u}_n^0,0,\dots,0]^\dag $, with
the only nonzero element at the $n$ position, and
$\mu_n=\sqrt{\int_{-T/2}^{T/2}(\dot{u}_n^0)^2\d t}$.

Deriving Eq.~(\ref{eq:gklein}) with respect to $\ee$ at $\ee=0$ we get:
\begin{equation}
\big(\N_0\,\frac{\partial u}{\partial \ee}\big)_n\,\equiv
 \, m_n\, \ddot{u}_{n,\ee}\,+\,V_n''(u^0_n)\,u_{n,\ee}\,=
 \,-\,\sum_m\,
\frac{\partial^2 \,W(u^0)}{\partial u_n\,\partial u_m}
\label{eq:guee}
\end{equation}

Following exactly the same procedure as in section~\ref{sec:multi} we
get the perturbation matrix $Q$, in extended form, with non diagonal
elements for indexes corresponding to excited oscillators:
\begin{equation}
Q_{n\,m}\,=\,\la n|\partial^2_{u\,u}W(u^0) | m \ra
\,=\,\frac{1}{\mu_n\, \mu_m}\int_{-T/2}^{T/2}\,\dot{u}_n
\frac{\partial^2 W(u^0) }{\partial u_n\,\partial u_m}\,
\dot{u}_m\,\d\,t\,, \quad n\neq m\,\,
 \label{eq:qnm}
\end{equation}
and diagonal elements
\begin{equation}
Q_{n\,n}=-\sum_{\forall\, m\neq n}\,\frac{\mu_m}{\mu_n}\,Q_{n\,m} \,.
 \label{eq:qnn}
\end{equation}
If there are $p$ oscillators at rest and $N-p$ excited
oscillators,  the extended matrix $Q$ has $p$ zero columns and
rows corresponding to the indexes of the oscillators at rest. The
reduced form of $Q$ is the $N-p$ squared matrix, equal to $Q$ but
stripped of those zero rows and columns. As a result we can
establish the theorem:
\begin{theorem}[Generalized MST]
Given a generalized Klein--Gordon system (\ref{eq:ghamiltonian}), a
specific multibreather solution at zero coupling $\{u^0_n\}$, determined
by a suitable set of codes, $u$ the corresponding solution at low
coupling, $\{\lambda\}_{i=1}^{N-p}$ the eigenvalues of the reduced
matrix $Q$, with only one zero, and being the wells corresponding to the
$N-p$ specific $u^0_n$, of the same type, hard or soft, then:\\ The
solution $u$ is stable if:
\\a) The on--site potentials are soft and there is not any positive
value in $\{\ee\,\lambda_i\}_{i=1}^{N-p}$.
\\b)The on--site potentials are hard and there is not any negative
value in $\{\ee\,\lambda_i\}_{i=1}^{N-p}$
\end{theorem}

This theorem includes both theorems~\ref{th:MST} and \ref{th:non--symmetric}.
Certainly, it is of not so
straightforward applicability, although for certain cases there are only
a few matrix elements $Q_{n\,m}$ to calculate. As an example, if there
are just two types of sites, the on--site potentials are symmetric, the
coupling is linear and we consider only time--reversible multibreathers,
there are only two different matrix elements to calculate.

\section{Applications}
\label{sec:applications}
\subsection{Two--site breathers}
\label{sec:twosite}
 A trivial example is the theorem about the
stability of the two--site breather found in Ref.~\cite{M97} page 69. Here
we generalize it to non--symmetric on--site potentials.
The coupling matrix corresponding to periodic boundary conditions and of
the standard (spring--type) attractive form, has elements
$C_{n\,m}\,=\,2\,\delta_{n\,m}-\delta_{|n-m|,1}$ and
$C_{N\,1}\,=\,C_{1\,N}\,=\,-1$, that is:
\begin{equation}
C\,=\,\left[ \begin{array}{rrrrrr} \;2&\;-1&0&\dots&\;0&\;-1\\
  -1 &2&-1&0&\dots&0\\
  \dots&\dots&\dots&\dots&\dots&\dots\\
  -1&0&\dots&0&-1&2
  \end{array}
 \right]\,. \label{eq:bigc}
\end{equation}

If the two oscillator are in phase, the perturbation matrix is
 $$
 Q\,=\,
 \left[
 \begin{array}{rr} \;1 &\; -1\\-1 &1 \end{array}\right]\,,
 $$
 with eigenvalues $(\lambda_1,\lambda_2)=(0,2)$, the first one corresponds
 to the phase mode, the second to the antisymmetric eigenmode.
 Thus,
one of the two bands tangent to the axis $E=0$ at zero coupling will
remain there, and the other will move upwards. If the on--site potential
is soft, i.e., with positive curvature, a intersection/tangent point
will be lost, and the two site breather is unstable. For the hard
on--site potential, the curvature of the bands is negative, and when the
band moved upwards, the intersection points will be kept, and the system
is be stable. If $\ee<0$ the signs of the eigenvalues $\ee\,E_i$ are
reversed and so are the conclusions.

For the  out--of--phase, two-site breather, with code
$\sigma=[1,-1]$,  the perturbation matrix is:
 $$Q\,=\,
 \left[
 \begin{array}{rr} \;-\gamma &\; \gamma\\\gamma &-\gamma \end{array}\right]\,,
 $$
with eigenvalues $(\lambda_1,\lambda_2)=(0,-2\gamma)$. The conclusions above are reversed,
being the instability mode, when appropriate, the symmetric one. Note
that the boundary conditions in $C$ have no effect.
\begin{theorem}[Two--site breathers]
\label{th:twosite} Consider time--reversible, two--site breathers in an homogeneous
Klein-Gordon system with attractive linear coupling, then:\\
 a)~The two-site breathers with codes $\pm\,[1,1]$ and soft on--site
 potential are unstable and with $\pm\,[-1,1]$ are stable.\\
 b)~The two-site breathers with  code $\pm\,[1,1]$ and hard on--site
 potential are stable and with $\pm\,[-1,1]$ are unstable.\\
 c)~If the coupling is repulsive, the conclusions are reversed.
\end{theorem}
\subsection{Three--site breathers}
Let us renumber the sites $1,2,3$ and suppose that the code of the site
in the middle is $\sigma_2,=\,+1$ as a reference, i.e., $\s\,=\,[\s_1,
1, \s_3]$. We define $\alpha_i=1$ if $\s_i=1$ and $\alpha_i=-\gamma$ if
$\s_i=-1$, for $i=1,2$.
Using $C$ from Eq.~(\ref{eq:bigc}) the perturbation matrix
is:
\begin{equation}
Q\,=\,\left[ \begin{array}{rrr}
  \; \alpha_1 \;&\;-\alpha_1 \; &\;0\\
   -\alpha_1 \;& \;\alpha_1+\alpha_3 \;&\; -\alpha_3\\
   0\;&\; -\alpha_3\; &\; \alpha_3
  \end{array}
 \right]\,. \label{eq:w3}
\end{equation}
Its nonzero eigenvalues are $\lambda_{\pm}\,=\, \alpha_1\,+\,
\alpha_3\,\pm\,\sqrt{(\alpha_1\,+\,\alpha_3)^2\,-\,3\,\alpha_1\,\alpha_3}$.
In order
to have both eigenvalues with the same sign, we need that  $\alpha_1\,
\alpha_3>0$, i.e., either $\alpha_3\,=\,\alpha_1\,=\,1$,
 or $\alpha_3\,=\,\alpha_1\,=\,-\gamma$.
That is, if
 $\s_1=\s_3=1$, $\lambda_\pm=1,3>0$, if $\s_1=s_3=-\gamma$,
  $\lambda_\pm=-\gamma,-3\gamma<0$, and
 if $\s_1\,\s_2=-1$, then $\lambda_\pm$ have different signs.   Therefore:

\begin{theorem}[Three--site breathers]
\label{th:threesite} Consider time--reversible, three--site breathers in an
homogeneous K-G system with nearest--neighbor, linear coupling, attractive for
$\ee>0$. Then:\\
  a)~The three--site breathers with codes
$\sigma\,=\,\pm [-1,1,-1]$ are stable if $\ee>0$ ($\ee<0$)and the
on--site potential is soft (hard).\\
 b)~The three--site breathers
with code $\sigma\,=\,\pm [1,1,1]$ are stable if $\ee>0$ $(\ee<0)$ and
the on--site potential is hard (soft).\\
 c)~All other three-site breathers are unstable for any sign of $\ee$
and type of on--site potential.
\end{theorem}

We can compare the predictions of the values given by the theory with
the numerically calculated values of the intersection points of the
bands with the axis $\theta=0$. As an example, the numerically
calculated eigenvalues and the predictions given by
theorems~\ref{th:MST} and \ref{th:non--symmetric} for the three--site
breather with code $[-1,1,-1]$ are plotted in
Figure~\ref{fig:morseasym}. The very good accuracy is evident.
%%%%%%%%%%%%%%%%%%%%%%%figure3%%%%%%%%%%%%%%%%%%%%%%%%%
\begin{figure}
\begin{center}
    \includegraphics[width=0.75\textwidth]{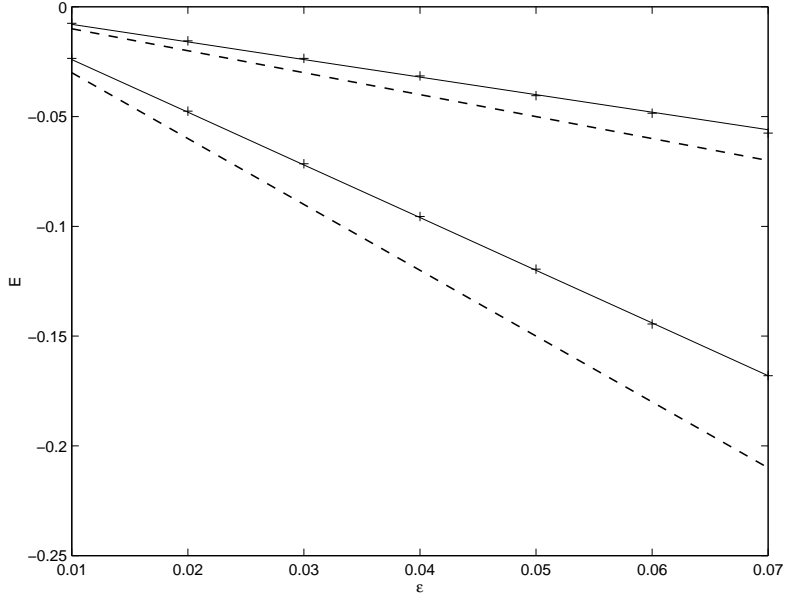}
\end{center}
\caption{Comparison between the eigenvalues $E$ of the Newton
operator obtained numerically (points), predicted by the symmetric MST,
theorem~\ref{th:MST}, (dashed lines) and by the non--symmetric MST,
theorem~\ref{th:non--symmetric}, (continuous line), which is the right
one to be used. Morse potential, code $[-1,1,-1]$, frequency
$\wb=\gamma=0.8$.}
\label{fig:morseasym}
\end{figure}

\subsection{A model with long--range interaction}
An example of system with long range interaction is the twist
model used in Ref.~\cite{SAPR02}. The dynamical equations are:
\begin{equation}
    \ddot{u}_n\,+\,V'(u_n)+ \ee \sum_{m=n-N/2}^{n+N/2} \frac{\cos[\theta_{tw} (n-m)]}
{|n-m|^3}\;u_m=0\,
 \label{eq:dyntwist}
\end{equation}
$V$ being the Morse potential, and $\theta_{tw}$ the angle between
two dipole moments corresponding to neighboring base pairs in a
simplified model of DNA. The coupling matrix elements are:
 \begin{equation}
 C_{n\,m}=  \frac{\cos[\theta_{tw} (n-m)]} {|n-m|^3}\; ,
\end{equation}
 Some of the results, also checked numerically and with good
accuracy with respect to the eigenvalues $E$, given by
theorem~\ref{th:MST}, in spite of $V$ being non--symmetric are:
\begin{center}
\begin{tabular}{|c|c|c|} \hline Code& $\theta_{tw}=0$ &
$\theta_{tw}=\pi$ \\ \hline \hline 11 & Stable & Unstable
\\\hline \hline 1 -1 & Unstable & Stable
\\ \hline 101 & Stable & Stable \\ \hline 111 & Stable & Unstable \\
\hline 1-1 1 & Unstable & Stable\\ \hline 11-1 & Unstable &
Stable\\ \hline
\end{tabular} \end{center}

As an example of the effect of the symmetry coefficient $\gamma$ in
theorem~\ref{th:non--symmetric}, we can consider the code $[1,-1,1]$ and
plot the Newton eigenvalues $E$ as a function of $\theta_{tw}$. The
perturbation matrix is:
\begin{equation}Q=\, \left[ \begin{array}{ccc}
\;\gamma a-b\;\;&\;\; -\gamma a & b\\ -\gamma a\;\;&\;\; 2 \gamma a& -\gamma a\\
b\;\;&\;\;-\gamma a\;\;&\;\; \gamma a-b
\end{array} \right]\;,
\end{equation}
with $a=\cos(\theta_{tw})$ and $b=\cos(2\theta_{tw})/8$.
Figure~\ref{fig:twist} shows the dependence of the eigenvalues
$\lambda$ on the twist angle $\theta_{tw}$ for three values of
the symmetry coefficient $\gamma$. It can be seen that only values
of $\gamma$ very far from $1$ lead to a change on the stability
prediction.

%%%%%%%%%%%%%%%%%%%%%%%%%%%figure4%%%%%%%%%%%%%%%%%%
\begin{figure}
\begin{center}
    \includegraphics[width=0.75\textwidth]{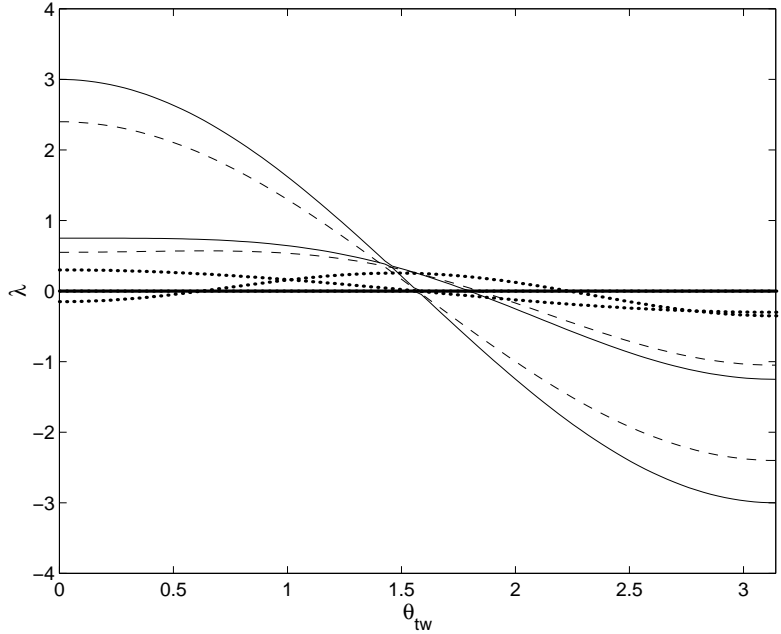}
\end{center}
\caption{Eigenvalues $\lambda$ of the perturbation matrix $Q$ for
a three-site breather with code $[-1,1,-1]$ on the twist model.
Continuous line: $\gamma=\wb=1$; dashed line: $\gamma=\wb=0.8$;
and dotted line: $\gamma=\wb=0.1$. Only values of $\gamma$ very
far from $1$ lead to a different stability prediction. }
\label{fig:twist}
\end{figure}

\section{Multibreathers, phonobreathers and dark breathers. Parity instabilities.}
\label{sec:dark}

For larger, time--reversible multibreathers in homogeneous Klein--Gordon systems
with nearest--neighbor, linear coupling, it is trivial to obtain the stability
properties of the two different solutions with wave number $q=0$, all
the oscillators in phase, or $q=\pi$, neighboring oscillators out of
phase.  Consider any number $N'$ of contiguous oscillators in phase
in a system coupled by the matrix $C$ in Eq.~(\ref{eq:bigc}). The
$N'\,\times\, N'$ perturbation matrix is given by:
 \begin{equation}
Q\,=\,\left[ \begin{array}{rrrrrr} \;1&\;-1&\;0&\;\dots&\;0&\;0\\
  -1 &2&-1&0&\dots&0\\
  \dots&\dots&\dots&\dots&\dots&\dots\\
  0&0&\dots&-1&2&-1\\
  0&0&\dots&0&-1&1
  \end{array}
 \right]\,. \label{eq:qmulti}
\end{equation}
The characteristic equation for $Q\,x=\lambda\,x$ corresponds to the
following equations:
\begin{eqnarray}
 x_1\,-\,x_2\,&=&\,\lambda\,x_1\,,\nonumber
 \\ 2x_n\,-\,x_{n+1}\,-\,x_{n-1}\,&=&\,\lambda\,x_n, \quad \forall \,n\,\neq\,1,2\,,\nonumber\\
 x_{N'}\,-\,x_{N'-1}\,&=&\,\lambda\,x_{N'}\,.
 \label{eq:eq:qequations}
\end{eqnarray}
Substitution of the trial solution $x_n=\cos(q\,n-\phi_{q})$, leads to
$\phi_q=q/2$, $q=m\,\pi/N'$, with $m=0,\dots,N'-1$, and
$\lambda_m=4\,\sin^2(m\,\pi/(2\,N'))$, i.e., positive eigenvalues
without a degenerate zero. In case all the oscillators of the group are
out of phase, $Q\rightarrow-\gamma Q$ and $\lambda_m\rightarrow-\gamma \lambda_m$.
Therefore:

\begin{theorem}
\label{th:multi}
If $\ee>0$, the time--reversible, in-phase, multibreathers are
unstable (stable) with soft (hard) on--site potential. For any
out--of--phase multibreathers, and $\ee<0$ the conclusions are reversed.
\end{theorem}
Although, we have no mathematical proof, according to numerical
calculations of the eigenvalues of the perturbation matrices
corresponding to groups with different codes, the numbers of negative
and positive eigenvalues are equal to the numbers of $-1$ and $+1$ in
$\{\sigma_n\,\sigma_{n+1}\}_{n=1}^{N'-1}$. Therefore being unstable for
any on--site potential.

Note that if there are several non--contiguous multibreathers,
the perturbation matrix if composed of independent blocks of perturbation submatrices of
the same type, each one with a zero eigenvalue by construction as seen in
subsection~\ref{sec:global}.
The stability of such a system cannot be
determined by the MST. The only possible conclusion is that if there are eigenvalues
of different sign this system would be unstable.

{\em Phonobreathers} are multibreathers without oscillators at rest. For
time--reversible phonobreathers in a system either with free--ends, or
fixed-ends boundary conditions, the perturbation matrix $Q$ is the same
as in Eq.~(\ref{eq:qmulti}). If the boundary conditions are periodic,
the only change is that $Q_{1\,1}=Q_{N,N}=2$, $Q_{N\,1}=Q_{1,N}=-1$, and
the eigenvalues are the same. That is, the mode $q=0$ with attractive
coupling and hard potential is stable. Changing the hardness, the type
of coupling, or the mode, changes each time the stability. In this way,
we reproduce the results in Refs.~\cite{DDP97,MJKA02} as a consequence
of modulational instability and using the DNLS approximation,
respectively, and in theorem~9 in Ref.~\cite{A97}, based in the
properties of the action.

{\em Parity instabilities} are a interesting effect for phonobreathers
with periodic boundary conditions. It is a consequence of the parity of
the total number of sites $N$. If, for example, $V$ is soft, the
coupling is attractive, and $N$ is even, all the eigenvalues of the
perturbation matrix are negative, i.e., the system is stable. But if $N$
is odd, there is an isolated positive eigenvalue, bringing about the
instability. There is a clear physical origin of this phenomenon, as the
boundaries are equivalent to an inhomogeneity. According to our theory
it is also obvious: we have, in fact two oscillators at the ends coupled
and in phase, which gives rise to a positive eigenvalue.

{\em Dark breathers} are multibreathers with only one or a few
oscillators at rest. Their stability depends on the characteristics of
the contiguous groups of excited oscillators, as a consequence of
theorem~\ref{th:multi}. We have to be careful about the number of
particles, for example the $\pi$ mode with code around the dark site
$[\dots,1,-1,0,1,-1,\dots]$, needs an odd number of sites to avoid
having two in--phase oscillators at the ends, and a parity instability.
The one with $[\dots,-1,1,0,1,-1,\dots]$ needs an even number of sites.
If the system has free of fixed periodic conditions, the perturbation matrix
is decoupled in two and the stability is undefined as commented above.

We have compared the predictions of the instabilities and the values of
the intersection points of the bands with the axis $\theta=0$, in many
cases, as for example in Ref.~\cite{AACR02}, with very good results. The
question of up to which values of the coupling parameter $\ee$ the
theory is valid, is however unsolved. Generically speaking values up to
$\ee=0.1$ (compared to the rest frequency $\omega_0=1$), are safe, but
there are some exceptions. In the same reference, for the dark breather
 hard
on--site potential and attractive coupling, a mode not predicted by the
MST gives rise to an instability as soon as $\ee=0.022$.

As the width of the phonon band can be calculated, and the width of the
background is given by the eigenvalues of the perturbation matrices, it
is also possible to predict the values of the $\ee$, for which
oscillatory and subharmonic instabilities occur, but we do not extend
further here.
\section{Summary and conclusions}
We have developed using degenerate perturbation theory a method for
obtaining the stability properties of multibreathers of any code at low
coupling. The method is synthesized in three different versions of a
theorem, referred to as the {\em multibreather stability theorem} or MST
for short. The two simplest versions correspond to time--reversible
multibreathers, with linear coupling and homogeneous on--site
potentials. If the on--site potential is symmetric or we consider only
in--phase multibreathers, it involves only some simple modifications of
the coupling matrix. If none of those conditions are fulfilled then it
involves the analytical or numerical calculations of a magnitude
$\gamma$, called the symmetry coefficient, as its value is $1$ for
symmetric on--site potentials. For soft potentials and values of the
frequency until about one half of the rest frequency, a good
approximation is the multibreather frequency, result which we
demonstrate is exact for the Morse potential. For hard potentials its
value its close to unity. The generalized version of the MST is of not
so straightforward applicability, although its complexity depends on
the characteristics of the system and multibreather considered.

We give some examples of application of the method and compare it with
numerical results. The systems considered are two and three--site
breathers, with nearest neighbor or long--range interactions,
multibreathers, phonobreathers and dark breathers. A parity instability
 can appear in finite systems depending on the parity of the number of
oscillators. All these examples are interesting in themselves, but also
 illustrate the use of
the MST, and show that it provides a powerful method
of easy applicability to determine multibreathers stability.

Some other applications under study are multibreathers in 2D and 3D
systems, prediction of the oscillatory and subharmonic instabilities and
the peculiarities of some disordered systems

\begin{ack}

This work has been supported by the European Union under the RTN 
project, LOCNET, HPRN--CT--1999--00163. J Cuevas acknowledges an 
FPDI grant from `La Junta de Andaluc\'{\i}a'. \end{ack}

\appendix
\section{Calculation of the symmetry parameter $\gamma$ for a
Morse potential} \label{sec:gammamorse}

In this appendix, we calculate the value of the symmetry parameter
$\gamma$ in theorem~\ref{th:non--symmetric} for the particular case of a
Morse potential. The choice of this kind of potential relies in the fact
that the expressions of the orbits are easy to manage. We will proof
that, for this potential, $\gamma=\wb$.

Let us suppose an isolated oscillator submitted to a Morse potential.
The energy of the system is:
\begin{equation}
\E=\frac{1}{2}\dot{x}^2 + \frac{1}{2}(\exp(-x)-1)^2\,.
\label{eq:hmorse}
\end{equation}
We look for solutions with period $T=2\pi/\wb$ that are
time--reversible, i.e, with $\dot{x}(0)=0$, or, in other words, at $t=0$
the system is at a turning point. There are two of them, obtained from
the equation above: $x_{1}=-\log(1+\sqrt{2\,\E})<0$ and
$x_{2}=-\log(1-\sqrt{2\,\E})>0$. If we consider the solution with
$x(0)=x_1$, t(x) is given by:
\begin{equation} \label{eq:t1}
    t(x)=\int_{x_1}^{x}\frac{\mathrm{d}x}{\sqrt{2(\E-(\exp(-x)-1)^2/2)}} \;,
\end{equation}
which leads to:
\begin{equation}
    t(x)=\frac{1}{\sqrt{1-2\,\E}}  \left(\frac{\pi}{2}-\arcsin\big[\frac{1}{\sqrt{2\,\E}}
    \;\big(1+\frac{2\,\E -1}{\exp(-x)}\;\big)\big]\right) \,.
\label{eq:t2}
\end{equation}
By substitution of $t=T/2=\pi/\wb$ and $x=x_2$ we obtain
$\wb=\sqrt{1-2\,\E}$.

The inversion of Eq.~(\ref{eq:t1}) leads to:
\begin{equation}\label{xt}
    x(t)=\log\frac{1-\sqrt{1-\wb^2}\cos\wt}{\wb^2}\,.
\end{equation}
To calculate $\gamma$, we need the Fourier coefficients of $x(t)$. If
the solution $x(t)$ is expressed as:
\begin{equation}
    x(t)=z_0+2\sum_{k=1}^{\infty}z_k\cos(k\,\wb\, t)\;,
\end{equation}
the coefficients $\{z_k\}_{k=0}^{\infty}$ are given by:
\begin{equation}
    z_k=\frac{1}{T}\int_{-T/2}^{T/2}x(t)\cos(k\,\wb\,t)\d t\,.
\end{equation}
The calculation of these integrals is straightforward and leads to:
\begin{equation}\label{zk}
     z_0=\log\frac{\displaystyle 1+\wb}{\displaystyle 2\wb^2} \quad;\quad
    z_k=-\,\frac{\displaystyle 1}{\displaystyle n}\Big(\;\frac{\displaystyle  1-\wb}
    {\displaystyle  \sqrt{1-\wb^2}}\;\Big)^k\,.
\end{equation}
The parameter $\gamma$ is given by:
\begin{equation}
    \gamma=-\frac{\sum_{k\ge1}(-1)^kk^2z_k^2}{\sum_{k\ge1}k^2 z_k^2}.
\end{equation}
By substitution of the Fourier coefficients in Eq.~(\ref{zk}), we obtain
\begin{equation}\label{gammaz}
    \gamma=-\frac{\sum_{k\ge1}(-r)^k}{\sum_{k\ge1}r^k}\,\quad \textrm{with}\quad
    r=\frac{1-\wb}{1+\wb}\,.
\end{equation}
As $|r|<1$, $\gamma$ can be easily calculated using the expressions for
the sum of an infinite geometric progression:
\begin{equation}
     \sum_{k\ge1}r^k=\frac{\displaystyle r}{\displaystyle 1-r} \quad,\quad
    \sum_{k\ge1}(-r)^k=-\frac{\displaystyle r}{\displaystyle 1+r}
\end{equation}
leading to $\gamma=\wb$, as we wanted to proof.

\section{Curvature of the bands} 
%appendix
\label{sec:curvature} In this 
section we demonstrate a key point used in our theory: the fact 
that the band corresponding to an isolated excited oscillators 
has negative curvature if the on site potential is hard and 
positive if it is soft. Unfortunately, the demonstration is quite long if we
want it to be self--consistent. A
similar one can be found in Ref.~\cite{Cre98} and  different ones 
 in Ref.~\cite{M97,KA99}.

Let us consider the dynamical equation of an isolated oscillator 
with Hamiltonian $H=1/2\,p^2+V(u)$, with $u$ a single--value, real function
of time and $p=\dot{u}$: 
 \begin{equation} 
 \label{eqb:dyn}
 \ddot{u}+V'(u)=0\,
 \end{equation}
 The characteristic equation for the Newton operator $\N$ 
 corresponding to a given periodic solution of Eq.~(\ref{eqb:dyn})is
 \begin{equation}
 \label{eqb:newton}
 \N\xi\equiv \ddot{\xi} +V''(u)\xi=E\xi \,,
 \end{equation}
 where $\xi$ is a $\mathcal{C}^2$ function of $t$. For each eigenvalue $E$ there are only 
 2 independent solutions, determined, for example, by the values of $\xi(0)$ and 
$\dot{\xi}(0)$. We can  obtain them for $E=0$, by deriving 
Eq.~(\ref{eqb:dyn}) with respect to $t$ and with respect to $\wb$:
\begin{equation} 
 \dddot{u}+V''(u)\dot{u}=0 \Leftrightarrow \N \dot{u}=0 \quad ;\quad
  \ddot{u}_{w}+V''(u) u_{w}=0 \Leftrightarrow \N u_{w}=0 \,.
 \end{equation}
As we have a single oscillator we can suppose that $u(t)$ is time 
symmetric with a suitable origin of time, i.e., $\dot{u}$ is a 
time-antisymmetric function of $t$ with the same period 
$T=2\,\pi/\wb$ as $u(t)$: $\dot{u}(t+T)=\dot{u}(t)$,  
$\dot{u}(0)=\dot{u}(T)=0$. The properties of $u_w=du/d\wb$, can 
be also obtained. We can write $u$ as a cosine Fourier series 
 $u(t)=\sum_{k=0}^{\infty}c_k \cos(\wt)$, with the coefficients $c_k$ depending on
 $\wb$. Therefore, ${du}/{d\wb} = \sum_{k=0}^{\infty}(dc_k/d\wb) \cos(\wt) +
 \sum_{k=0}^{\infty}-k\,t\sin(\wt)$, or:
 \begin{equation}
 \label{eqb:uw}
 u_w\equiv\frac{du}{d\wb}=\gamma(t)+\frac{t}{\wb}\,\dot{u}(t)\,
 \end{equation}    
$\gamma(t)$ being a time symmetric, periodic function of time 
with the same period $T$ as $u$. This functions constitute a base 
of the eigenspace corresponding to $E=0$.

On the other hand, $\N$ (\ref{eqb:newton}) is a periodic operator because
 it depends
on a periodic function $u$. Its eigenfunctions are 
given by the Bloch theorem: 
 \begin{equation} \label{eqb:xi}
 \xi(\theta,t)=\chi(\ii\theta,t)\exp(\theta t/T), 
 \end{equation} with 
$\chi$ a T--periodic function of time and $\theta=\theta(E)$. If 
we consider only values of $E$ for which $\theta$ is real, the 
set of points $(\theta,E(\theta)$ constitutes a band. If $\xi$ is 
a solution of $\N\,\xi=E\xi$, the fact that $\N$ is a real operator
implies that $\xi^*$ 
is another solution, i.e., $(-\theta,E)$ also belongs to the band
and this is symmetric with respect to $\theta$ and, therefore,
 $\d E/\d \theta (0)=0$. As $\N$ is time--symmetric 
$\xi(\theta,-t)$ is another, and has to be proportional to $\xi^*$ as  
there can be only two independent solutions.   

The Floquet matrix $F_E$ is defined as: 
 \begin{equation}
 \left[ \begin{array}{c} \xi(T)\\ \dot{\xi}(T)\end{array}\right]=
 F_E \left[ \begin{array}{c} \xi(0)\\ \dot{\xi}(0)\end{array}\right] \,.
 \end{equation}
The eigenvalues $\{\lambda_l\}$ of $F_E$ are the Floquet 
multipliers, if we write them as $\lambda_l=\exp{\ii\theta_l}$, 
with $\theta$ real or complex, $\ii \theta_l$ are the Floquet 
exponents, and $\theta_l$, the Floquet arguments. An 
eigenfunction with $\theta_l$ real is bounded, if $\theta_l=1$
the eigenfunction has period $T$. The condition for linear 
stability of the solution $u$ is that all the Floquet arguments
of $F_0$ are real. 

It is easy to check some properties. The Bloch functions 
Eq.~(\ref{eqb:xi}) are eigenfunctions of $F_E$ with Floquet 
exponent $\theta$, therefore $F_E$ is diagonalizable (over 
$\mathcal{C}$) with Floquet multipliers $\exp(\pm \ii\theta)$ and 
Floquet arguments $\pm \theta$. However, at $E=0$, there is only 
one eigenfunction of $F_0$, $\dot{u}$, the other independent 
function of the subspace $\N \,\xi=0$, $u_w$, it is not an 
$F_0$--eigenfunction. That is $F_0$ has a degenerate Floquet 
multiplier $1$, or Floquet argument $0$ and can be transformed 
into a Jordan block. Another consequence of the latter is that 
$(\theta,E)=(0,0)$ belongs to the band. We need to calculate the 
curvature of the band at this point.

Let us come back to the Bloch functions $\xi(\theta,t)$ in 
Eq.~({\ref{eqb:xi}) and  $\xi^*(\theta,t)$ 
corresponding to a pair of symmetric points $(\pm\theta,E)$  of 
the band, and suppose that $\theta\rightarrow 0$ (and then $E 
\rightarrow 0$). At $\theta=0$ the two functions collide in
$\xi(0,t)=\chi(0,t)=\chi^*(0,t)$, therefore $\xi(0,t)$ is a real, 
$T$--periodic function. But we already now that at $E=0$, there 
is a $T$--periodic function solution of $\N\,\xi=0$, $\dot{u}$. 
Therefore, $\xi(0,t)=\dot{u}$ ($\N$ is lineal, so we can adjust 
the norm of $\xi$ in order that the latter equation is 
fulfilled).      

We can reobtain the missing function that spans $\ker(\N)$. If we 
derive Eq.~(\ref{eqb:newton}) with respect to $\theta$, we obtain 
$\N\,\xi_\theta=E_\theta\xi + E\xi_\theta$ and at $E=0$ 
($E_\theta(0)=0$) we obtain: $\N\,\xi_\theta(0,t)=0$. Deriving 
$\xi(\theta,t)$ in Eq.~(\ref{eqb:xi}) we get: 
 \begin{equation}
 \label{eqb:xitheta}
 \xi_\theta = \chi_\theta(\theta,t) \exp(\ii\theta/T)+\ii\frac{t}{T}
 \chi(\theta,t)\exp(\ii\theta/T) =
 \chi_\theta \exp(\ii\theta/T)+\ii\frac{t}{T}\xi \,.
 \end{equation}
At $\theta=0$, $\xi=\dot{u}$ and $\xi_\theta$ has to be 
proportional at $u_w$, which we know is the other independent 
function. Therefore
 \begin{eqnarray}
 \nonumber
 \xi_\theta(0,t)=\chi_\theta +\ii\frac{t}{T}\,\dot{u}=
 \ii\frac{\wb}{T}\,\left( \frac{T}{\ii\wb}\,\chi_\theta+\frac{t}{\wb}\,\dot{u}\right)
 =\ii\frac{\wb^2}{2\pi}\,
 \left( -\frac{2\pi}{\wb^2}\,\chi_\theta+\frac{t}{\wb}\,\dot{u}\right)\,.
 \end{eqnarray}
Comparing with Eq.~(\ref{eqb:uw}) we obtain 
 \begin{equation}\label{eqb:xithuw}
   \xi_\theta(0,t)=\ii \frac{\wb^2}{2\pi}\,u_w .
 \end{equation}
 
 The following step is to express the curvature $E_{\theta\theta}$ of the band
  $E(\theta)$ at $(\theta,E)=(0,0)$ in terms of
 $u$ and its derivatives. Let us derive Eq.~(\ref{eqb:newton}) twice with respect to $\theta$, 
 we obtain:
 \begin{equation}\label{eqb:ntheta}
 \N\,\xi_{\theta\theta}=E_{\theta \theta}\,\xi + 
 2\,E_{\theta}\,\xi_{\theta}+E\,\xi_{\theta \theta}
\Rightarrow \N\,\xi_{\theta\theta}=E_{\theta\theta}\,\dot{u} \quad
\mathrm{at}\,\, \theta=0\,.
 \end{equation}
 Multiplying to the right by $\dot{u}$ and integrating a period we get
 \begin{equation}\label{eqb:eiint}
 E_{\theta\theta}\,\int_{t=0}^T \dot{u}^2\d t = 
 \int_{t=0}^T \dot{u}\N\,\xi_{\theta\theta} \d t \,.
 \end{equation}
 For any pair of functions $f(t)$ and $g(t)$
 \begin{eqnarray}
 \int_{t=0}^T f\N\,g&=&
 \int_{t=0}^T (f\,\ddot{g}+f\,V''\,g)\d t = 
 [f\,\dot{g}]_{t=0}^T-[\dot{f}\,g]_{t=0}^T + \int_{t=0}^T (g\ddot{f}+g\,V''\,g)\,d t
 \nonumber \\&=&
  [f\,\dot{g}]_{t=0}^T-[\dot{f}\,g]_{t=0}^T +\int_{t=0}^T g\,\N\,f\d t \,.
 \nonumber
 \end{eqnarray}
 Applying the latter equation:
 \begin{eqnarray}
\int_{t=0}^T \dot{u}\N\,\xi_{\theta\theta} \d t =
  [\dot{u}\,\dot{\xi}_{\theta\theta}]_{t=0}^T
  -[\ddot{u}\,\xi_{\theta\theta}]_{t=0}^T +
  \int_{t=0}^T (\xi_{\theta\theta}\,\N\,\dot{u})\d t \,.
 \end{eqnarray}
 Using that $\N\,\dot{u}=0$, $\dot{u}(0)=\dot{u}(T)=0$ and 
 $\ddot{u}(0)=\ddot{u}(T)$ the previous equation leads to
 \begin{equation}
 \int_{t=0}^T \dot{u}\,\N\,\xi_{\theta\theta} \d t =
 -\ddot{u}(T)\left(\xi_{\theta\theta}(0,T)-\xi_{\theta\theta}(0,0)\right) \,.
 \end{equation}
 To calculate the last difference, let us derive Eq.~(\ref{eqb:xitheta}) with respect
 to $\theta$. We obtain
 \begin{eqnarray}
 \xi_{\theta\theta}&=&\chi_{\theta\theta}\exp(\ii\theta t/T)+ 
 \ii\frac{t}{T}\,\chi_{\theta}\exp(\ii \theta t/T)+\ii\frac{t}{T}\,\xi_\theta
  \nonumber\\&=& \chi_{\theta\theta}\exp(\ii\theta t/T) +
 \ii\frac{t}{T}\,( \xi_{\theta}-\ii\frac{t}{T}\,\xi)+
 \ii \frac{t}{T}\,\xi_{\theta}\nonumber\\
 &=&
 \chi_{\theta\theta}\exp(\ii\theta t/T) +
 2\ii\frac{t}{T}\,\xi_{\theta} + \frac{t^2}{T^2}\xi \nonumber\\
 &\Rightarrow&\quad \xi_{\theta\theta}=
 \chi_{\theta\theta} +
 2\ii\frac{t}{T}\,\xi_{\theta} + \frac{t^2}{T^2}\dot{u}
 \quad \mathrm{at}\,\,\,\theta=0 \nonumber\,.
 \end{eqnarray}
 $\chi$ and therefore $\chi_{\theta\theta}$ are $T$--periodic functions, therefore
 $\xi_{\theta\theta}(0,0)=\chi_{\theta\theta}(0,0)=\chi_{\theta\theta}(0,T)$ and
 $
 \xi_{\theta\theta}(0,T)-\xi_{\theta\theta}(0,0)=2\ii\,\xi_{\theta}(0,T)\,.
  $ Then
  \begin{equation}
  \int_{t=0}^T \dot{u}\,\N\,\xi_{\theta\theta} \d t=
  -\ddot{u}\,(2\ii\xi_{\theta}(0,T))=\frac{\partial H}{\partial u(T)}\,
  2\ii \,(\ii\frac{\wb^2}{2\pi}\,u_w(T))=
  -\frac{\wb^2}{\pi}\frac{\d H}{\d \wb}\nonumber \,.   
    \end{equation}
Substituting in Eq.~(\ref{eqb:eiint}), and naming the action 
$I=\int_{t=0}^T\dot{u}^2\,\d t$ we finally obtain at 
$\theta=0$:
 \begin{equation}
 E_{\theta\theta}(0)=-\frac{\wb^2}{\pi I}\,\frac{\d H}{\d\wb}\,.
 \end{equation} 
$I$ is always positive, therefore, if the potential is hard ($\d H/\d\wb>0$),
the curvature  $E_{\theta\theta}$ is negative and the band is tangent to 
the axis $E=0$ from below, if it is soft ($\d H/\d\wb<0$) the curvature is
positive and the band is tangent form above, as we wanted to demonstrate.

Note that this demonstration is also valid for the whole
coupled system with the obvious changes: $u\rightarrow [u_1,\dots,u_N]^\dag$;
$V(u)\rightarrow V(u)+\ee W(u)$, $W(u)$ being the coupling potential;
$V'(u)\rightarrow [\partial V/\partial u_1,\dots,\partial V/\partial u_N]^\dag$;
 $V''(u)$ the matrix with elements $\partial^2 V/\partial u_n \partial u_m$ 
 and analogously. We also 
 need the additional constraint that the solution $u(t)$ is chosen
 time--symmetric and that the pair of eigenvalues with $\theta=0$ is isolated.
 This means that the band tangent at $(\theta,E)=(0,0)$ continues having
the same curvature properties depending on the hardness/softness of the potential

%\bibliography{todas}

\begin{thebibliography}{10}

\bibitem{MA94}
RS~MacKay and S~Aubry.
\newblock Proof of existence of breathers for time-reversible or
  \mbox{Hamiltonian} networks of weakly coupled oscillators.
\newblock {\em \mbox{Nonlinearity}}, 7:1623--1643, 1994.

\bibitem{MA95}
JL~Mar{\'{\i}}n and S~Aubry.
\newblock Breathers in nonlinear lattices: Numerical methods based on the
  anti--integrability concept.
\newblock In L~V\'{a}zquez, L~Streit, and VM~P\'{e}rez-Garc\'{\i}a, editors,
  {\em Nonlinear {K}lein--{G}ordon and Schr\"{o}dinger Systems: Theory and
  Applications}, pages 317--323. World Scientific, Singapore and Philadelphia,
  1995.

\bibitem{A97}
S~Aubry.
\newblock Breathers in nonlinear lattices: Existence, linear stability and
  quantization.
\newblock {\em \mbox{Physica D}}, 103:201--250, 1997.

\bibitem{FW98}
S~Flach and CR~Willis.
\newblock Discrete breathers.
\newblock {\em Physics Reports}, 295:181--264, 1998.

\bibitem{M97}
JL~Mar{\'{\i}}n.
\newblock {\em Intrinsic Localised Modes in Nonlinear Lattices.}
\newblock PhD thesis, University of Zaragoza, Department of Condensed Matter,
  June 1997.

\bibitem{MJKA02}
AM~Morgante, M~Johansson, G~Kopidakis, and S~Aubry.
\newblock Standing waves instabilities in a chain of nonlinear coupled
  oscillators.
\newblock {\em Phys. D}, 162:53, 2002.

\bibitem{DDP97}
I.~Daumont, T.~Dauxois, and M.~Peyrard.
\newblock Modulational instability: fisrt step towards energy localization in
  nonlinear lattices.
\newblock {\em Nonlinearity}, 10:617--630, 1997.

\bibitem{G95}
D~Griffiths.
\newblock {\em Introduction to Quantum Mechanics}.
\newblock Prentice Hall, New Jersey, {USA}, 1995.

\bibitem{SAPR02}
B~S\'anchez-Rey, JFR Archilla, F~Palmero, and FR~Romero.
\newblock Breathers in a system with helicity and dipole interaction.
\newblock {\em Phys. Rev. E}, 66:017601--017604, 2002.

\bibitem{AACR02}
A~Alvarez, JFR Archilla, J~Cuevas, and FR~Romero.
\newblock Dark breathers in {K}lein-{G}ordon lattices. band analysis of their
  stability properties.
\newblock {\em New Journal of Physics}, 4:72.1--72.19, 2002.

\bibitem{Cre98}
T~Cretegny.
\newblock {\em Dynamique collective et localisation de l'\'energie dans les
  reseaux non-lin\'{e}aires}.
\newblock PhD thesis, \'Ecole Normale Sup\'erieure de Lyon, 1998.

\bibitem{KA99}
G~Kopidakis and S~Aubry.
\newblock Intraband discrete breathers in disordered nonlinear systems {I}:
  Delocalization.
\newblock {\em \mbox{Physica D}}, 130:155--186, 1999.

\end{thebibliography}
%\bibliographystyle{unsrt}
%\end{document}
\newcommand{\noopsort}[1]{} \newcommand{\printfirst}[2]{#1}
  \newcommand{\singleletter}[1]{#1} \newcommand{\switchargs}[2]{#2#1}
  \ifx\undefined\allcaps\def\allcaps#1{#1}\fi\ifx\undefined\allcaps\def\allcap%
s#1{#1}\fi

\end{document}